# Ancient Australian Rocks and the Search for Life on Mars

A. J. Brown, Plancius Research
C. E. Viviano, JHU Applied Physics Laboratory
T. A. Goudge, University of Texas at Austin

Abstract: We discuss the results of a remote sensing study that has revealed new details about an important rock unit dominated by two minerals that can be associated with volcanism (olivine) and life (carbonate). The study, which used a new analysis technique on CRISM data, identified a region where no carbonates or clays are present, only large grain size olivine. This discovery shines new light on the formation and history of the olivine-carbonate rock within Jezero crater that will be explored by the Mars 2020 rover.

# 1. Introduction

Figure 1 shows the location of Jezero crater, the landing site of the $2.42 billion Mars 2020 rover. One of the most compelling reasons Jezero was chosen as the landing site is the presence of an ancient geological unit that contains the primitive mineral olivine. The extent of this olivine unit is shown in light green, as recently mapped by Kremer et al., (2019) across the broader Nili Fossae region.

Olivine was first identified at Nili Fossae in 2003 by the Thermal Emission Spectrometer (Hoefen et al., 2003) on NASA's Mars Global Surveyor spacecraft and subsequently further mapped by the Thermal Emission Imaging System (THEMIS) instrument onboard the Mars Odyssey orbiter (Koeppen and Hamilton, 1998). Excitement about the unit escalated when carbonate minerals were found to be associated with the olivine unit (Ehlmann et al., 2008). The carbonate was discovered using the Compact Reconnaissance Imaging Spectrometer for Mars (CRISM) visible-near infrared hyperspectral instrument on NASA's Mars Reconnaissance Orbiter spacecraft.

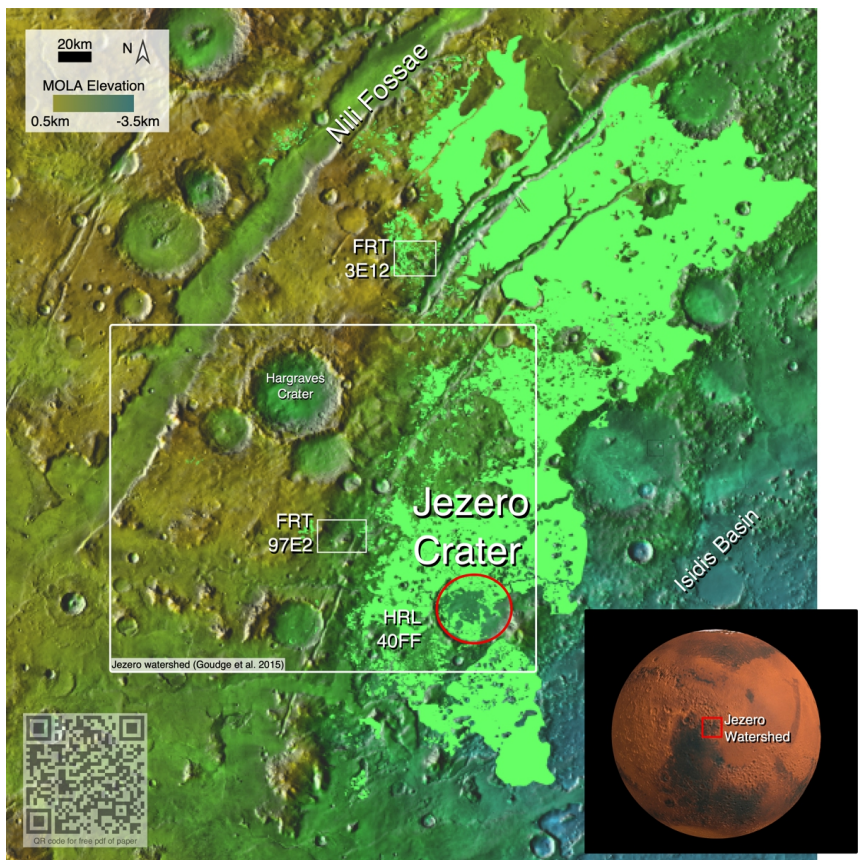

Fig. 1. The extent of the olivine-carbonate lithology showing the Nili Fossae study region as mapped by Kremer et al. (2019). Jezero crater is shown in the lower part of the map and the location on Mars in the inset. Credit: Authors

## 2. Background

### 2.1 Early Habitability of Earth and Mars

Jezero crater once held a lake, fed by two inlet river valleys, each of which has a delta deposit identified at the valley mouth (Goudge et al., 2015). Deltas form here on Earth when a river channel discharges into a lake or ocean basin. The water, which had been travelling swiftly in the channel is able to entrain sediments as it flows. Once the water reaches a lake or ocean basin it is suddenly bought to a halt and can no longer hold the sediment, and so it is deposited at the mouth of the channel.

The olivine-carbonate unit underlies the Jezero deltas and is incorporated into carbonate deposits around the edge of the crater recently hypothesized to have formed in standing lake waters (Horgan et al., 2020). Horgan et al., (2020) have even suggested that stromatolites could be present in the "marginal carbonates" around the edge of the crater. Stromatolites form when photosynthetic microbes trap sand and sediment to create successive layers, which can then be trapped in the rock record (Fig. 2). Stromatolites have been discovered in the Strelley Pool Chert in the Pilbara region of Western Australia, and constitute some of the best evidence of an early terrestrial biome at 3.43 Ga (Walter et al., 1980, Buick 1990, Van Kranendonk et al., 2008).

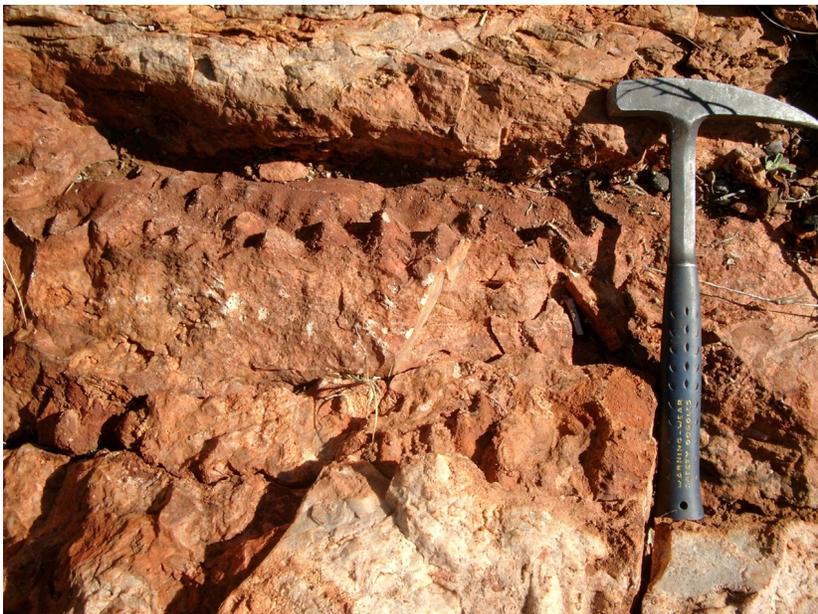

Fig. 2 Conical Stromatolites from the 3.43 Ga Strelley Pool Chert in Western Australia, Fig 3g of Brown et al., (2004).

The similarities in age between the Strelley Pool stromatolites and the olivine-carbonate lithology are striking. Since life on Earth had gotten started prior to 3.5 billion years ago, and was likely in a shallow marine environment (Allwood et al., 2004) could life on Mars have also gotten started quickly, and taken advantage of a relatively clement surface environment during the special time in Mars' history before the planet was purged of its liquid water? Is there more that we could learn about the first environments of terrestrial life that would inform our search for life on Mars?

## 2.2 Serpentinization and Komatiite

A decade ago, a paper by Brown et al., (2010) suggested that there was a considerable amount of spectral similarity between the olivine-carbonate lithology on Mars and an ultramafic dunite in the Pilbara in Western Australia. The dunite was sourced in the 3.47 billion year old Mount Ada basalt, which is at the bottom of the ultramafic-mafic-felsic-chert sequence that is capped by the stromatolites in the 3.35 billion year old Strelley Pool Chert. In brief, they made the point that a process of hydrothermal serpenitinization was one way to form the clays and carbonate that are both observed with olivine at the same time (Bishop et al. 2008, 2011; Roush et al. 2015). They posited that the phyllosilicate might be talc, in analogy with Archean talc-carbonate dunites, an idea which was also investigated by Viviano et al., (2013) and McSween et al., (2014). In particular, the Mount Ada Basalt dunite is an example of a komatiite rock, which have been suggested as good analogs for Martian lava flows (Figure 3).

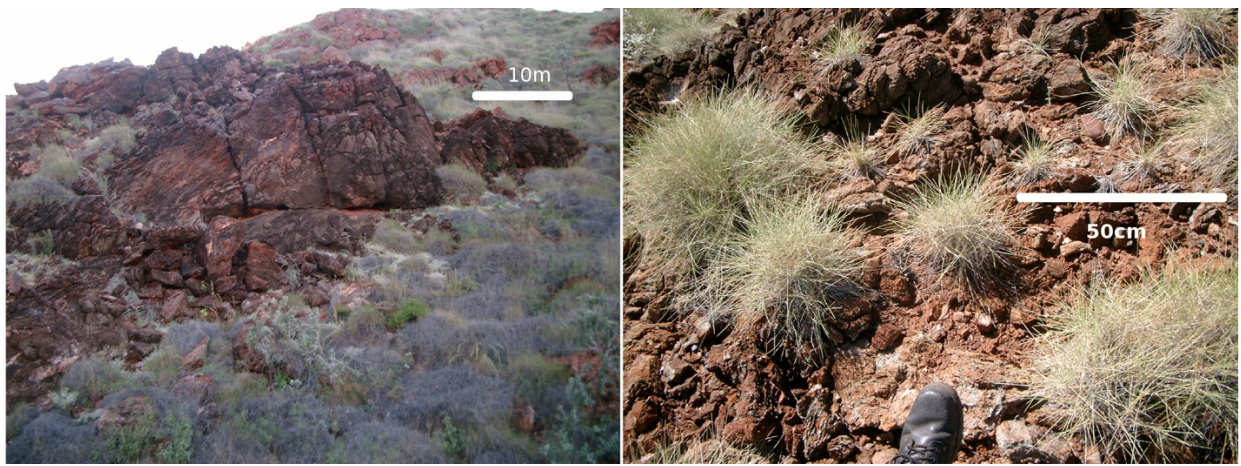

Fig. 3 Clay-carbonate altered basalt outcrop in the Pilbara region, Fig. 1 of Brown et al. (2010).

Komatiite lavas form when high-temperature (~1400-1600 °C), low viscosity (0.1-1 Pa s), mantle derived, ultramafic lavas are extruded and flow turbulently at the surface. Komatiites are found almost exclusively in Archean shield areas due to the higher heat of the Earth's mantle during that period (Campbell et al., 1989). Treiman (2005) has convincingly demonstrated that the Nakhalite meteorites have a similar texture to komatiite cumulate layers in the Canadian Archean. Komatiite lavas have also previously been proposed as possible analogs for Martian rocks on geochemical (Baird and Clark, 1981), morphological (Reyes and Christensen, 1994; Rampey and Harvey, 2012) and microtextural (Treiman, 2005) grounds. We know that large grained olivine crystals is a typical (though not unique) signature of komatiite lava, but - is there anything else to be learned about the physical properties of the olivine-carbonate layer?

## 3. Results

The recent study of Brown et al. (2020) reported that within the olivine-carbonate lithology, there is band saturation of the olivine 1 μm band complex. Band saturation occurs when the grain size of the scattering crystal is large enough that photons entering the grain never make it out. Variability of this saturation effect enabled them to place limits on the grain size of olivine controlling the band shape. They also used an Asymmetric Gaussian to fit the centroid of the 1μm spectral band and produced maps in three key regions of Nili Fossae (Brown, 2006).

They showed, via a spectral plot of 1 μm band centroid and asymmetry, that the presence of at least 500 microns grain size olivine is required, and at most 1 mm grains can be accommodated. Additionally, if the 1mm grain size is assumed, they showed that these can be best fitted with a medium (Fo #44-66) composition of olivine, rather than a high magnesium olivine, which had been previously suggested (e.g. Edwards and Ehlmann, 2015).

Finally, the Brown et al. (2020) study found that the region displaying the most redshifted olivine 1 μm bands (i.e., bands shifted to longer wavelengths) was located in a part of the olivine-carbonate lithology north of Jezero crater. Interestingly, they observe that these most redshifted olivine spectra display no carbonate or phyllosilicate bands. This suggests that this could be a least altered 'populaition zero' part of the lithology.

This study has shown where carbonates are *not* being formed, but – are there any other constraints on where the carbonates came from?

# 4. Discussion

## 4.1 Komatiite lavas

Komatiite lavas form when high-temperature (~1400-1600 °C), low viscosity (0.1-1 Pa), mantle derived, ultramafic lavas are extruded and flow turbulently at the surface. Komatiites are found almost exclusively in Archean shield areas due to the higher heat of the Earth's mantle during that period (Campbell et al., 1989). Treiman (2005) has convincingly demonstrated that the Nakhalite meteorites have a similar texture to komatiite cumulate layers in the Canadian Archean. Komatiite lavas have also previously been proposed as possible analogs for Martian rocks on geochemical (Baird and Clark, 1981), morphological (Reyes and Christensen, 1994; Rampey and Harvey, 2012) and microtextural (Treiman, 2005) grounds. The komatiite layer detected in the North Pole Dome was associated with talc-carbonate alteration which has been hypothesized to be the result of hydrothermal alteration (Brown et al., 2005, 2006, 2010). The main figure shows members of the Mars2020 investigating the talc-carbonate layer.

## 4.2 Origin of the Carbonate

The 2.42 billion dollar question is - where did all this carbonate come from? Some studies have advocated a model where a thick early $CO_2$ Noachian atmosphere reacted with the olivine unit to form the carbonate. The $CO_2$ atmosphere model does have to contend with the difficulty that there is no carbonate in the underlying basement unit, which would likely have been exposed to the atmosphere at the same time as the olivine unit. Instead, Brown et. al. (2020) entertained a suggestion that the carbonate may have been sourced from the Martian mantle at a time when $CO_2$ was being lost to the Martian mantle, according to a model of Grott et al. (2011). Edwards and Ehlmann (2015) earlier calculated the amount of $CO_2$ required to form the Nili Fossae carbonates as the equivalent of 0.25 mbar $P_{CO2}$. The mantle loss model of Grott et al. (2011) estimates that ~240mbar of $CO_2$ was outgassed during the period 4.1-3.7Ga, and at least 19 mbar in the 3.7-2Ga period (their Table 2). In fact, Grott et al. (2011) put forward a fast loss and slow loss model and which process actually took place is unclear. Mars 2020 is well placed to determine whether the $CO_2$ source is volcanic or pyroclastic, by looking for signs of bombs or lapilli. If Mars2020 is then able to collect samples of the 3.82 billion year old olivine-carbonate lithology for analysis on Earth, we may be able to better understand the history of $CO_2$ in the mantle of Mars using isotopic laboratory methods to determine the provenance of the carbonate.

## 4.3 The Thermal Inertia Problem

The thermal inertia of the entire surface of Mars has been mapped at highest resolution by the THEMIS instrument. "Thermal inertia" is a measure of the surface's response to being heated. If it responds quickly, the thermal inertia is low, this is usually true of fine grained materials like sand. If the thermal inertia is high, the surface's temperature responds slowly to heating, like large rocks or competent lava flows. The thermal inertia of the olivine-carbonate lithology has already been reported as relatively low (Rogers et al., 2018) and this was interpreted as being potentially at odds with the idea of the unit being a lava flow (Kremer et al. (2019) and Mandon et al. (2020)).

The recent study of Brown et al. (2020) was interested in the question of whether there was a correlation between the thermal inertia and the grain size of olivine. They did not find any reliable correlation, and posed this as "The Thermal Inertia Problem". They suggested a couple of reasons why this might be the case, but pointed out the problem will have to be studied in situ by the Mars 2020 rover.

In addition, as exemplified in the main figure and Figure 3, talc-carbonate serpentinized material is highly friable and has very low thermal inertia. This may constitute a clue to solving full thermal inertia problem that will be further explored by the Mars 2020 rover.

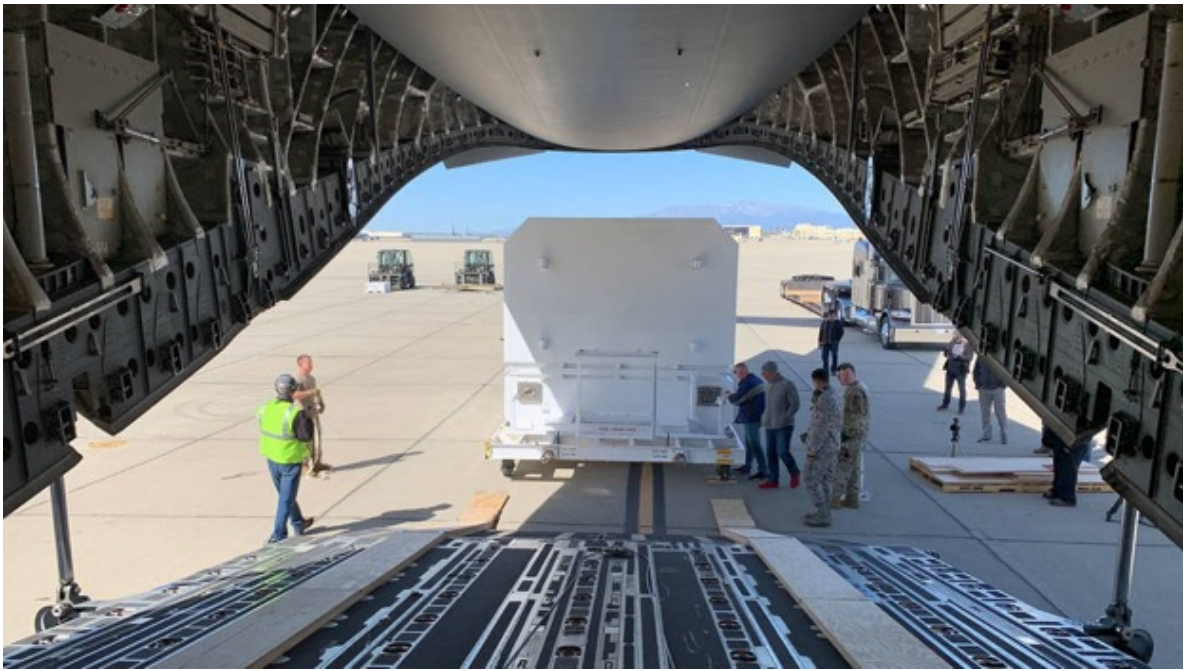

Fig. 4. Latest news from Mars 2020 – the rover arrives at Cape Canaveral 12 February 2020 for launch window opening July 17th. Credit: JPL/NASA

5. Conclusions

The Mars 2020 rover has just been delivered to Cape Canaveral in preparation for a July 17th launch on an Atlas 5 rocket, and will land at Jezero crater on 18 February 2021. After "seven minutes of terror" during its descent into the

Martian atmosphere, and a successful landing, Mars 2020 will be off to find the nearest olivine-carbonate outcrop or exposure.

Thanks to the remote sensing studies mentioned in this article, and many more studies before that, the Mars planetary science community is now better prepared than ever to land at a geologically diverse and compelling landing site in the most ancient rocks exposed on the surface. This landing site will allow us all to be transported back in time to when the Jezero delta system was being deposited, and water was running on the Martian surface (Williford et al., 2018).

It was a different time, indeed, and to best comprehend that Ancient Mars will require that we view it through the prism of our own astrobiological record.

## 6. Acknowledgements

This work was supported by the NASA Astrobiology Institute (Grant NNX15BB01A) and from NASA MDAP (Grant NNX16AJ48G). T. A. G. acknowledges support from the CRISM team through a subcontract from the Johns Hopkins University Applied Physics Lab.